\documentclass[a4paper,11pt]{article}
\usepackage{pos}

\title{Bright blazar flares with CTA}
 \ShortTitle{Bright blazar flares with CTA}

\author*[a]{M. Cerruti}
\author[b]{J. Finke}
\author[c]{G. Grolleron}
\author[c]{J.P. Lenain}
\author[d]{T. Hovatta}
\author[e]{M. Joshi}
\author[d]{E. Lindfors}
\author[f]{P. Morris}
\author[g]{M. Petropoulou}
\author[h]{P. Romano}
\author[h]{S. Vercellone}
\author[i]{M. Zacharias}

\affiliation[a]{Université Paris Cité, CNRS, Astroparticule et Cosmologie, F-75013 Paris, France}

\affiliation[b]{U.S. Naval Research Laboratory, Code 7653, 4555 Overlook Avenue SW, Washington, DC 20375-5352, USA}

\affiliation[c]{Laboratoire de Physique Nucléaire et des Hautes Energies (LPNHE), Sorbonne Université, Université Paris Cité, CNRS/IN2P3, F-75005, Paris, France}

\affiliation[d]{Finnish Centre for Astronomy with ESO (FINCA), University of Turku, Vesilinnantie 5, 20014 University of Turku, Finland
}

\affiliation[e]{Research Computing, Information Technology Services, Northeastern University, USA}

\affiliation[f]{Deutsches Elektronen-Synchrotron DESY, Platanenallee 6, D-15738 Zeuthen, Germany}

\affiliation[g]{Department of Physics, National and Kapodistrian University of Athens, University Campus, Zografos, GR 15783, Greece}

\affiliation[h]{INAF - Osservatorio Astronomico di Brera, Via Brera 28, 20121 Milano, Italy}

\affiliation[i]{Landessternwarte, Universität Heidelberg, Königstuhl, 69117, Heidelberg, Germany}

\onbehalf{for the CTA Consortium} 

\emailAdd{cerruti@apc.in2p3.fr}

\abstract{The TeV extragalactic sky is dominated by blazars, radio-loud active galactic nuclei with a relativistic jet pointing towards the Earth. Blazars show variability that can be quite exceptional both in terms of flux (orders of magnitude of brightening) and time (down to the minute timescale). This bright flaring activity contains key information on the physics of particle acceleration and photon production in the emitting region, as well as the structure and physical properties of the jet itself. The TeV band is accessed from the ground by Cherenkov telescopes that image the pair cascade triggered by the interaction of the gamma ray with the Earth's atmosphere. The Cherenkov Telescope Array (CTA) represents the upcoming generation of imaging atmospheric Cherenkov telescopes, with a significantly higher sensitivity and larger energy coverage with respect to current instruments. It will thus provide us with unprecedented statistics on blazar light-curves and spectra. In this contribution we present the results from realistic simulations of CTA observations of bright blazar flares, taking as input state-of-the-art numerical simulations of blazar emission models and including all relevant observational constraints. }

\ConferenceLogo{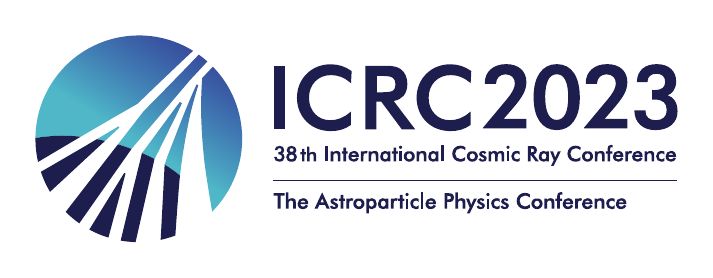}

\FullConference{%
38th International Cosmic Ray Conference (ICRC2023)\\
  26 July - 3 August, 2023\\
  Nagoya, Japan}

\begin{document}
\maketitle

\section{Introduction}
The current generation of Imaging Atmospheric Cherenkov Telescopes (IACTs), composed of the three arrays MAGIC, H.E.S.S., and VERITAS, has greatly increased our knowledge of the very-high-energy $\gamma$-ray (VHE, energies greater than 100 GeV) sky, bringing the number of known VHE sources from a dozen to about 250, in a bit less than twenty years of data taking \citep{tevcat}. The extra-galactic component of the VHE sky is dominated by active galactic nuclei (AGN), i.e. accreting super-massive black holes, of the blazar type. Within the unified AGN model, a blazar is a radio-loud AGN whose relativistic jet points in the direction of the observer. The relativistic boosting of the emission is what makes blazars particularly bright within the AGN population. They are characterized by non-thermal emission over a broad range of wavelengths, from radio up to VHE, a high degree of polarization in radio, optical, and X-rays, and they exhibit remarkable variability in both brightness (with significant increases spanning orders of magnitude) and time-scales (reaching as short as minute-scale variability). The rapid variability is of particular interest, because time changes in the emission encode important information about the physical properties of the emitting region, the emission processes at work in it, as well as the acceleration processes that are energizing the particles in the jet (leptons or hadrons) \citep{Blandford, CerrutiRev}. \\

The next generation IACT, the Cherenkov Telescope Array, CTA \citep{CTA}, is currently under construction. It will consist of two arrays, one in the Northern Hemisphere, on the Canary island of La Palma, close to the running MAGIC telescopes, and one in the Southern Hemisphere, at the Paranal Observatory in Chile. In order to maximize the scientific return of the instrument, the CTA Consortium is currently working on simulations of the expected outcomes of the observations. The work presented in this contribution is part of the preparation for the CTA AGN Key Science Project \citep{CTA}. What is discussed here represents a part of this larger effort, and focuses on the simulation of future CTA observations of blazar flares, concentrating on the study of rapid variability with a particular emphasis on the capability to reconstruct spectral variability. A complementary study (shown in these proceedings by \citet{Grolleron23}) focuses on the long-term variability. The preliminary results of this work have been presented in \citet{Cangemi23}. \\
\section{Simulations}

\begin{figure}[t!]
    \centering
    \includegraphics[width=7.5cm]{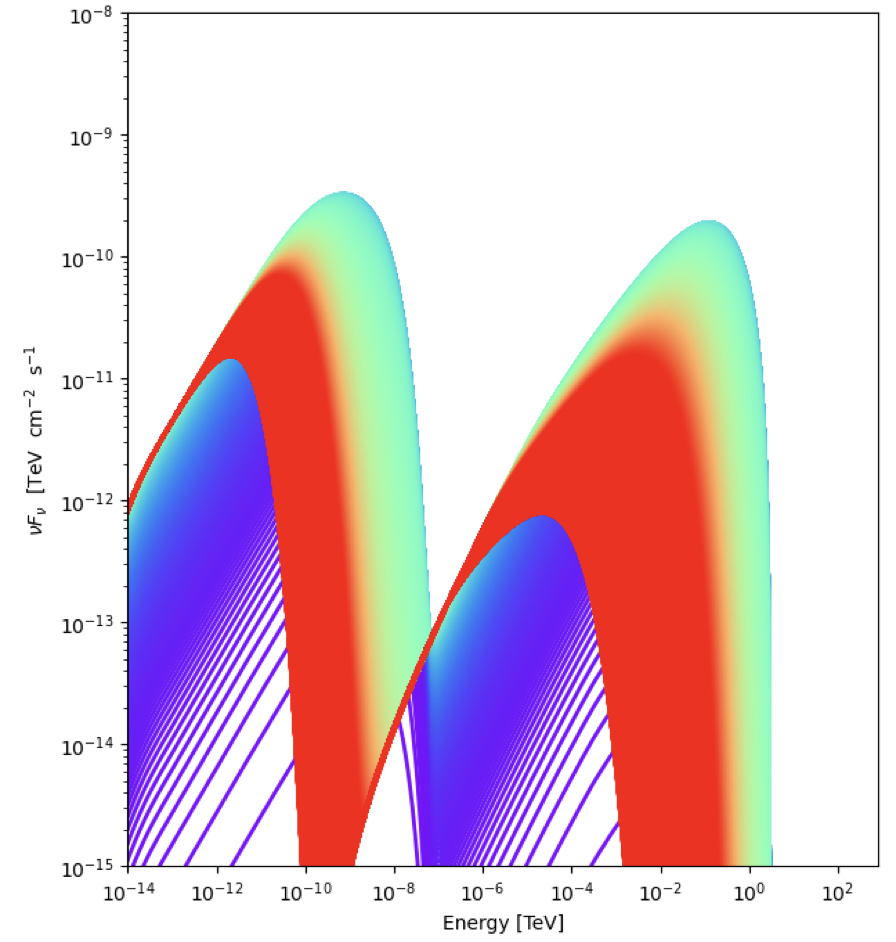}
    \includegraphics[width=7.5cm]{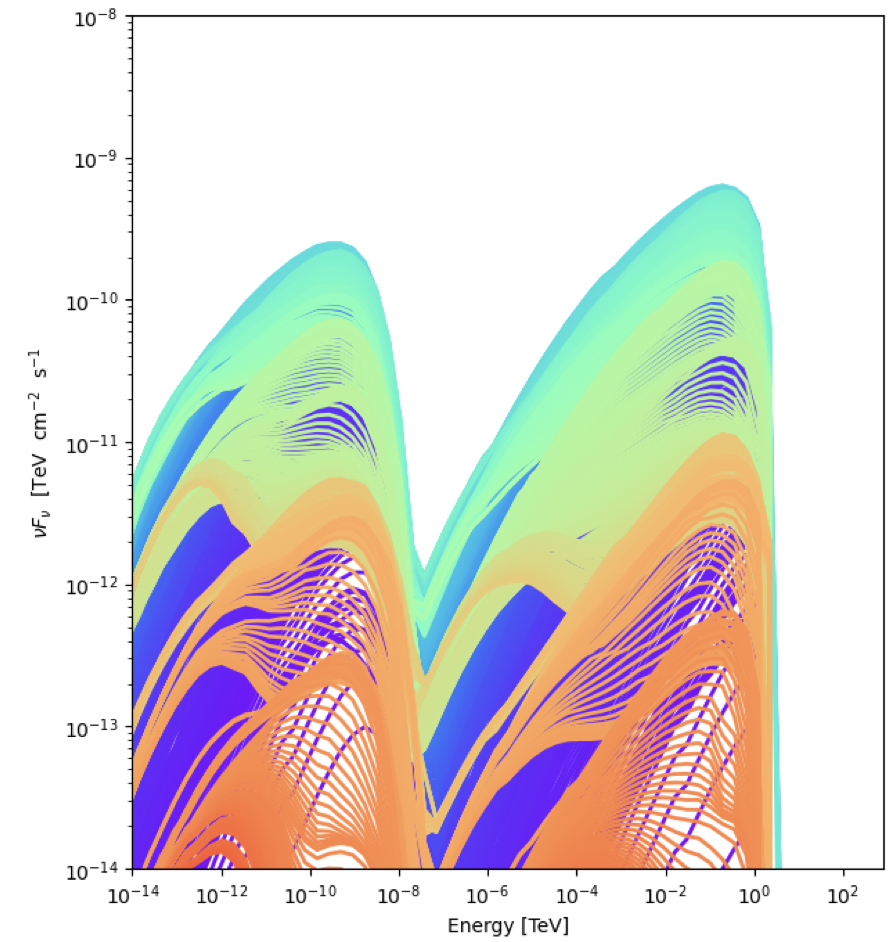}
    \caption{Theoretical SEDs provided as input for the CTA simulations. Left: model A; Right: model B (see text for details). The color code, from violet to red, represents the elapsed time.}
    \label{figureone}
\end{figure}

The first step of the simulation is to input theoretical models that have been developed to describe data from current observatories. In order to be as general as possible, we do not fit existing data, but we rather produce theoretical models that can approximately reproduce (in terms of flux and time variability) observed flares. In this contribution, we limit ourselves to two different models that approximately describe the variability observed in the well known VHE blazar Mrk 421. Input models are provided in the form of time-dependent spectral energy distributions, produced over a broad spectral range, from radio to VHE. The next step is then to simulate CTA observations: this is done using the \texttt{CTAAGNVAR} pipeline\footnote{\url{https://gitlab.cta-observatory.org/guillaume.grolleron/ctaagnvar}}, which is built upon the official CTA high-level analysis tool, Gammapy \citep{Gammapy}. \texttt{CTAAGNVAR} reads the theoretical AGN spectrum as input, and produces a simulated CTA observation including realistic observational constraints as outputs. As the zenith angle of the source will vary during an observing period, the software implements source tracking and selects the appropriate instrumental response functions (IRF). Once the CTA simulated spectra are produced, they are then fitted using phenomenological spectral functions in order of increasing complexity (i.e. a simple power-law, a log-parabola, a power-law with exponential cut-off; the more complex model is considered only if it improves the fit), as done by observers on real data. Absorption on the extragalactic background light is included when performing the fit. The best-fit model parameters can then be studied, in order to investigate the capability of CTA to reconstruct the input models and ultimately discriminate among them. In this contribution we focus on specific observational properties: the capability of CTA to reconstruct spectral variability and hysteresis whenever present in the input model. This is a very important feature, already detected in the X-ray band in blazars, predicted in the VHE band by some of the models, but as yet undetected in the VHE band \citep{VerMrK}. In the following, we show the results from two different theoretical inputs: a single-zone leptonic model in which the acceleration mechanism is not explicit, and electrons are assumed to be injected with a power-law shape and then cool down as they radiate (in the following, model A)\citep{Finke08}; and a flaring activity triggered by magnetic reconnection (in the following, model B)\citep{Christie19}. The input models are shown in Figure \ref{figureone}.

\section{Results}

\begin{figure}[t!]
    \centering
    \includegraphics[width=7.5cm]{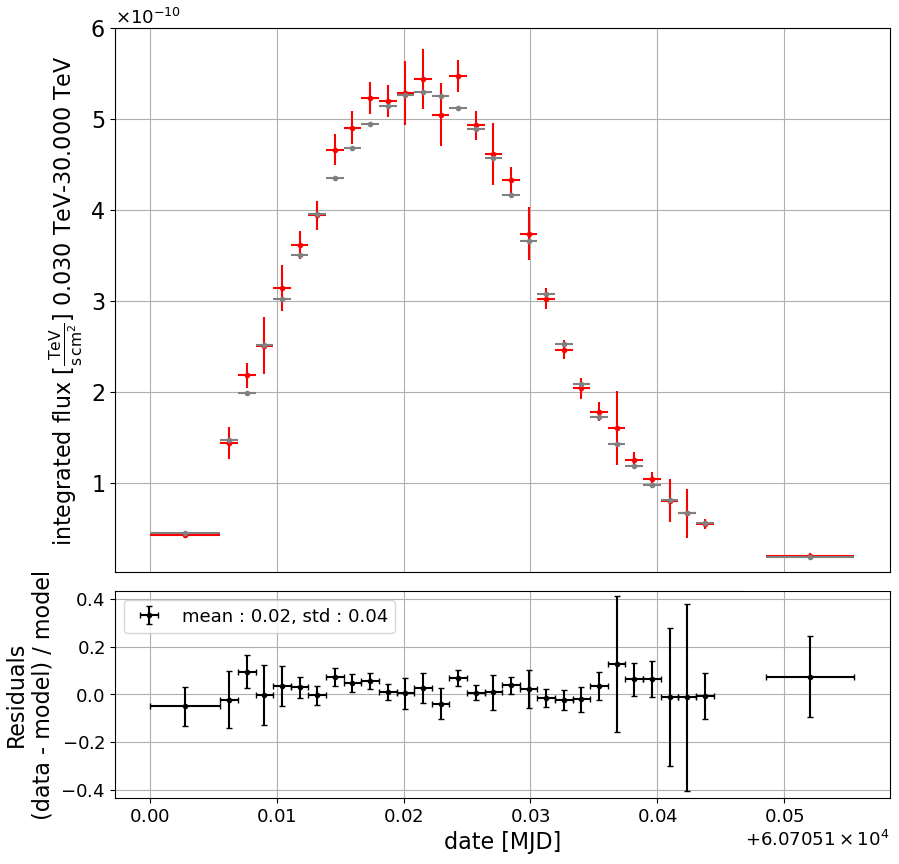}
    \includegraphics[width=7.5cm]{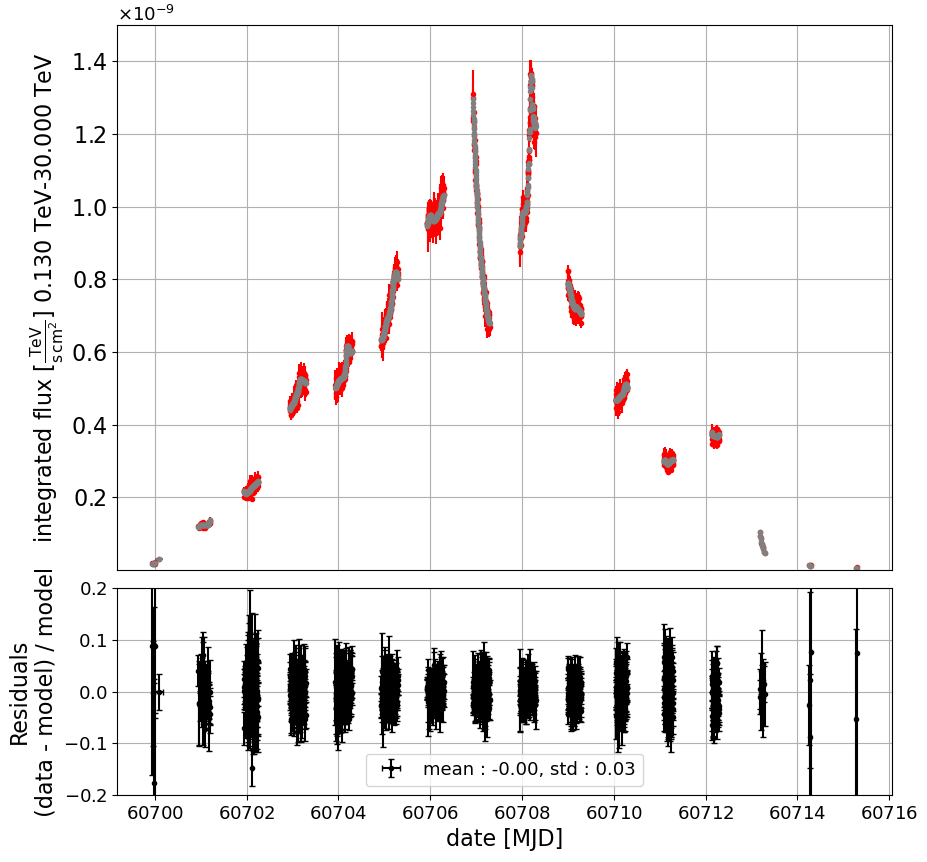}
    \caption{Simulated CTA light-curves, expressed as differential flux . Left: model A; Right: model B (see text for details).}
    \label{figuretwo}
\end{figure}

\begin{figure}[t!]
    \centering
    \includegraphics[width=7.5cm]{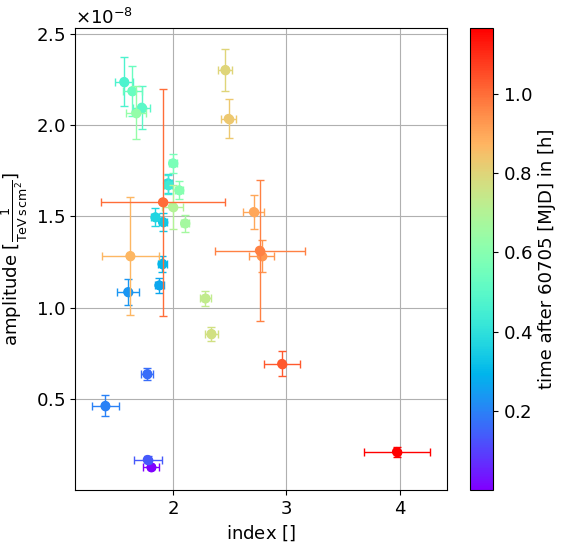}
    \includegraphics[width=7.5cm]{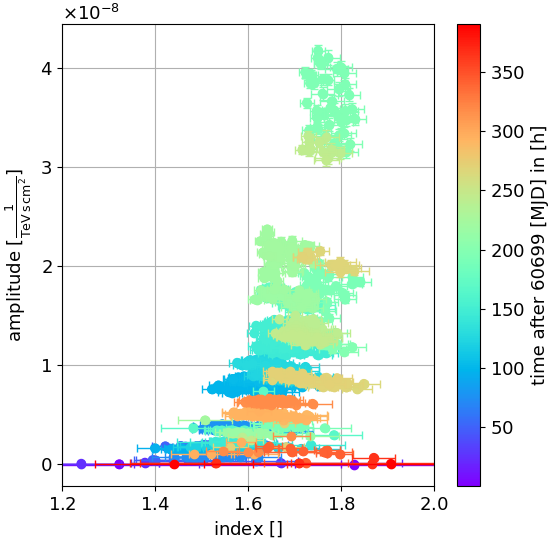}
    \caption{Differential flux versus best-fit power-law index. Left: model A; Right: model B (see text for details).}
    \label{figurethree}
\end{figure}

\begin{figure}[t!]
    \centering
    \includegraphics[width=7.5cm]{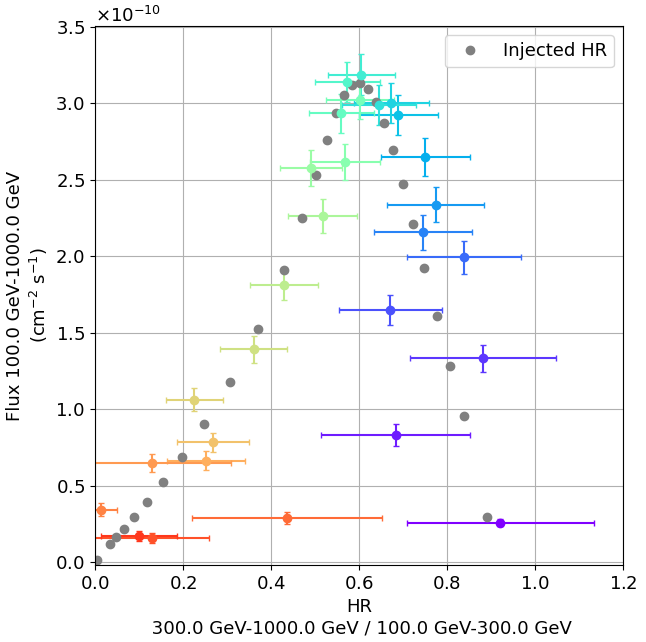}
    \includegraphics[width=7.5cm]{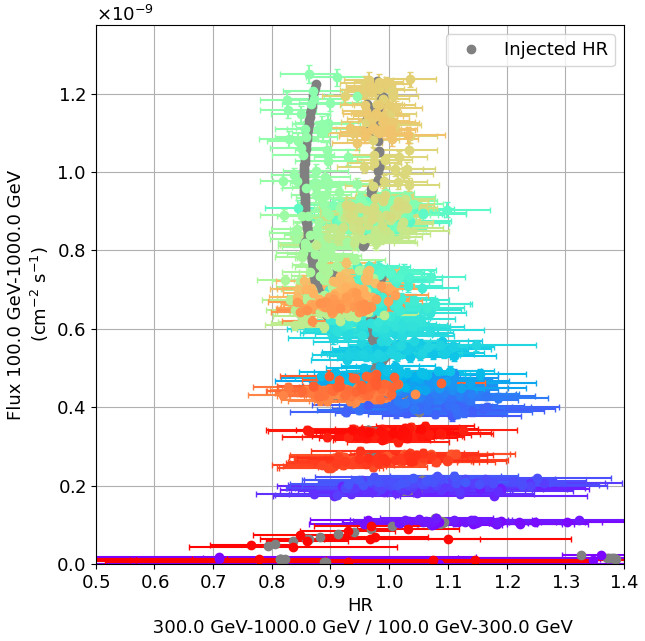}
    \caption{Hardness-ratio plot: flux in the low-energy band vs the hard/soft flux ratio. Left: model A; Right: model B (see text for details).}
    \label{figurefour}
\end{figure}

The results of the simulations are shown in Figures \ref{figuretwo} to \ref{figurefour}. In Figure \ref{figuretwo} we show simulated CTA light-curves (using the CTA North IRFs) for both models: model A represents a fast flare happening during a single observing night, while model B covers a larger data set of approximately two weeks, even though during the brightest nights fast intra-night variability can also be observed. In Figure \ref{figurethree} we show the results of a power-law fit to CTA data, plotted as amplitude vs photon index: these simulations indicate that model A has intrinsic spectral variability that can be detected by CTA; on the other hand model B shows weaker spectral variability in the CTA data. As an alternative to this visualization plot, we also produce two hardness-ratio plots, which is a common display tool in X-ray astronomy: in Figure \ref{figurefour} we show the evolution of the integral flux as a function of the hardness ratio between a high and low CTA energy band. Here as well we clearly observe the hysteresis cycle in the CTA data for model A.\\


\section{Conclusions}

CTA will provide unprecedented sensitivity in the VHE band, giving us access to  much increased statistical sample on blazar flares compared to current IACTs. In this contribution we have shown two simulated CTA light-curves on bright blazar flares, taking as input two different state-of-the-art numerical models. The preliminary results indicate that CTA might be able to detect, for the first time, hysteresis cycles in the VHE band, if they are indeed produced by the acceleration and radiative processes at work in the jet. This will give us a new observable to further constrain theoretical models. The results presented here are a small sub-set of the simulations that we are currently performing. \\

\section*{Acknowledgments}
Please see the full CTA acknowledgments at \url{https://www.cta-observatory.org/consortium_acknowledgments/}
\bibliography{ref}

\begin{thebibliography}{}
\expandafter\ifx\csname natexlab\endcsname\relax\def\natexlab#1{#1}\fi
\providecommand{\url}[1]{\href{#1}{#1}}
\providecommand{\dodoi}[1]{doi:~\href{http://doi.org/#1}{\nolinkurl{#1}}}
\providecommand{\doeprint}[1]{\href{http://ascl.net/#1}{\nolinkurl{http://ascl.net/#1}}}
\providecommand{\doarXiv}[1]{\href{https://arxiv.org/abs/#1}{\nolinkurl{https://arxiv.org/abs/#1}}}

\bibitem[{{Abeysekara} {et~al.}(2017){Abeysekara}, {Archambault}, {Archer},
  {Benbow}, {Bird}, {Buchovecky}, {Buckley}, {Bugaev}, {Cardenzana}, {Cerruti},
  {Chen}, {Ciupik}, {Connolly}, {Cui}, {Eisch}, {Falcone}, {Feng}, {Finley},
  {Fleischhack}, {Flinders}, {Fortson}, {Furniss}, {Griffin},
  {H{\r{a}}kansson}, {Hanna}, {Hervet}, {Holder}, {Humensky}, {H{\"u}tten},
  {Kaaret}, {Kar}, {Kertzman}, {Kieda}, {Krause}, {Kumar}, {Lang}, {Maier},
  {McArthur}, {McCann}, {Meagher}, {Moriarty}, {Mukherjee}, {Nieto}, {O'Brien},
  {Ong}, {Otte}, {Park}, {Pelassa}, {Pohl}, {Popkow}, {Pueschel}, {Ragan},
  {Reynolds}, {Richards}, {Roache}, {Sadeh}, {Santander}, {Sembroski},
  {Shahinyan}, {Staszak}, {Telezhinsky}, {Tucci}, {Tyler}, {Wakely},
  {Weinstein}, {Wilhelm}, {Williams}, {VERITAS Collaboration}, {Ahnen},
  {Ansoldi}, {Antonelli}, {Antoranz}, {Arcaro}, {Babic}, {Banerjee}, {Bangale},
  {Barres de Almeida}, {Barrio}, {Becerra Gonz{\'a}lez}, {Bednarek},
  {Bernardini}, {Berti}, {Biasuzzi}, {Biland}, {Blanch}, {Bonnefoy}, {Bonnoli},
  {Borracci}, {Bretz}, {Carosi}, {Carosi}, {Chatterjee}, {Colin}, {Colombo},
  {Contreras}, {Cortina}, {Covino}, {Cumani}, {Da Vela}, {Dazzi}, {De Angelis},
  {De Lotto}, {de O{\~n}a Wilhelmi}, {Di Pierro}, {Doert}, {Dom{\'\i}nguez},
  {Dominis Prester}, {Dorner}, {Doro}, {Einecke}, {Eisenacher Glawion},
  {Elsaesser}, {Engelkemeier}, {Fallah Ramazani}, {Fern{\'a}ndez-Barral},
  {Fidalgo}, {Fonseca}, {Font}, {Fruck}, {Galindo}, {Garc{\'\i}a L{\'o}pez},
  {Garczarczyk}, {Gaug}, {Giammaria}, {Godinovi{\'c}}, {Gora}, {Guberman},
  {Hadasch}, {Hahn}, {Hassan}, {Hayashida}, {Herrera}, {Hose}, {Hrupec},
  {Hughes}, {Idec}, {Kodani}, {Konno}, {Kubo}, {Kushida}, {Lelas}, {Lindfors},
  {Lombardi}, {Longo}, {L{\'o}pez}, {L{\'o}pez-Coto}, {Majumdar}, {Makariev},
  {Mallot}, {Maneva}, {Manganaro}, {Mannheim}, {Maraschi}, {Marcote},
  {Mariotti}, {Mart{\'\i}nez}, {Mazin}, {Menzel}, {Mirzoyan}, {Moralejo},
  {Moretti}, {Nakajima}, {Neustroev}, {Niedzwiecki}, {Nievas Rosillo},
  {Nilsson}, {Nishijima}, {Noda}, {Nogu{\'e}s}, {N{\"o}the}, {Paiano},
  {Palacio}, {Palatiello}, {Paneque}, {Paoletti}, {Paredes}, {Paredes-Fortuny},
  {Pedaletti}, {Peresano}, {Perri}, {Persic}, {Poutanen}, {Prada Moroni},
  {Prandini}, {Puljak}, {Garcia}, {Reichardt}, {Rhode}, {Rib{\'o}}, {Rico},
  {Saito}, {Satalecka}, {Schroeder}, {Schweizer}, {Shore}, {Sillanp{\"a}{\"a}},
  {Sitarek}, {Snidaric}, {Sobczynska}, {Stamerra}, {Strzys}, {Suri{\'c}},
  {Takalo}, {Tavecchio}, {Temnikov}, {Terzi{\'c}}, {Tescaro}, {Teshima},
  {Torres}, {Torres-Alb{\`a}}, {Toyama}, {Treves}, {Vanzo}, {Vazquez Acosta},
  {Vovk}, {Ward}, {Will}, {Wu}, {Zanin}, {MAGIC Collaboration}, {Hovatta}, {de
  la Calle Perez}, {Smith}, {Racero}, \& {Balokovi{\'c}}}]{VerMrK}
{Abeysekara}, A.~U., {Archambault}, S., {Archer}, A., {et~al.} 2017, ApJ, 834,
  2, \dodoi{10.3847/1538-4357/834/1/2}

\bibitem[{{Blandford} {et~al.}(2019){Blandford}, {Meier}, \&
  {Readhead}}]{Blandford}
{Blandford}, R., {Meier}, D., \& {Readhead}, A. 2019, ARA\&A, 57, 467,
  \dodoi{10.1146/annurev-astro-081817-051948}

\bibitem[{{Cangemi} {et~al.}(2023){Cangemi}, {Hovatta}, {Lindfors}, {Cerruti},
  {Becerra-Gonzalez}, {Biteau}, {Boisson}, {B{\"o}ttcher}, {de Gouveia Dal
  Pino}, {Dorner}, {Grolleron}, {Lenain}, {Manganaro}, {Max-Moerbeck},
  {Morris}, {Nilsson}, {Passos Reis}, {Romano}, {Sergijenko}, {Tavecchio},
  {Vercellone}, {Wagner}, \& {Zacharias}}]{Cangemi23}
{Cangemi}, F., {Hovatta}, T., {Lindfors}, E., {et~al.} 2023, arXiv e-prints,
  arXiv:2304.14208, \dodoi{10.48550/arXiv.2304.14208}

\bibitem[{{Cerruti}(2020)}]{CerrutiRev}
{Cerruti}, M. 2020, Galaxies, 8, 72, \dodoi{10.3390/galaxies8040072}

\bibitem[{{Cherenkov Telescope Array Consortium} {et~al.}(2019){Cherenkov
  Telescope Array Consortium}, {Acharya}, {Agudo}, {Al Samarai}, {Alfaro},
  {Alfaro}, {Alispach}, {Alves Batista}, {Amans}, {Amato}, {Ambrosi},
  {Antolini}, {Antonelli}, {Aramo}, {Araya}, {Armstrong}, {Arqueros},
  {Arrabito}, {Asano}, {Ashley}, {Backes}, {Balazs}, {Balbo}, {Ballester},
  {Ballet}, {Bamba}, {Barkov}, {Barres de Almeida}, {Barrio}, {Bastieri},
  {Becherini}, {Belfiore}, {Benbow}, {Berge}, {Bernardini}, {Bernardini},
  {Bernardos}, {Bernl{\"o}hr}, {Bertucci}, {Biasuzzi}, {Bigongiari}, {Biland},
  {Bissaldi}, {Biteau}, {Blanch}, {Blazek}, {Boisson}, {Bolmont}, {Bonanno},
  {Bonardi}, {Bonavolont{\`a}}, {Bonnoli}, {Bosnjak}, {B{\"o}ttcher},
  {Braiding}, {Bregeon}, {Brill}, {Brown}, {Brun}, {Brunetti}, {Buanes},
  {Buckley}, {Bugaev}, {B{\"u}hler}, {Bulgarelli}, {Bulik}, {Burton},
  {Burtovoi}, {Busetto}, {Canestrari}, {Capalbi}, {Capitanio}, {Caproni},
  {Caraveo}, {C{\'a}rdenas}, {Carlile}, {Carosi}, {Carqu{\'\i}n}, {Carr},
  {Casanova}, {Cascone}, {Catalani}, {Catalano}, {Cauz}, {Cerruti}, {Chadwick},
  {Chaty}, {Chaves}, {Chen}, {Chen}, {Chernyakova}, {Chikawa}, {Christov},
  {Chudoba}, {Cie{\'s}lar}, {Coco}, {Colafrancesco}, {Colin}, {Conforti},
  {Connaughton}, {Conrad}, {Contreras}, {Cortina}, {Costa}, {Costantini},
  {Cotter}, {Covino}, {Crocker}, {Cuadra}, {Cuevas}, {Cumani}, {D'A{\`\i}},
  {D'Ammando}, {D'Avanzo}, {D'Urso}, {Daniel}, {Davids}, {Dawson}, {Dazzi}, {De
  Angelis}, {de C{\'a}ssia dos Anjos}, {De Cesare}, {De Franco}, {de Gouveia
  Dal Pino}, {de la Calle}, {de los Reyes Lopez}, {De Lotto}, {De Luca}, {De
  Lucia}, {de Naurois}, {de O{\~n}a Wilhelmi}, {De Palma}, {De Persio}, {de
  Souza}, {Deil}, {Del Santo}, {Delgado}, {della Volpe}, {Di Girolamo}, {Di
  Pierro}, {Di Venere}, {D{\'\i}az}, {Dib}, {Diebold}, {Djannati-Ata{\"\i}},
  {Dom{\'\i}nguez}, {Dominis Prester}, {Dorner}, {Doro}, {Drass}, {Dravins},
  {Dubus}, {Dwarkadas}, {Ebr}, {Eckner}, {Egberts}, {Einecke}, {Ekoume},
  {Els{\"a}sser}, {Ernenwein}, {Espinoza}, {Evoli}, {Fairbairn},
  {Falceta-Goncalves}, {Falcone}, {Farnier}, {Fasola}, {Fedorova}, {Fegan},
  {Fernandez-Alonso}, {Fern{\'a}ndez-Barral}, {Ferrand}, {Fesquet},
  {Filipovic}, {Fioretti}, {Fontaine}, {Fornasa}, {Fortson}, {Freixas
  Coromina}, {Fruck}, {Fujita}, {Fukazawa}, {Funk}, {F{\"u}{\ss}ling},
  {Gabici}, {Gadola}, {Gallant}, {Garcia}, {Garcia L{\'o}pez}, {Garczarczyk},
  {Gaskins}, {Gasparetto}, {Gaug}, {Gerard}, {Giavitto}, {Giglietto}, {Giommi},
  {Giordano}, {Giro}, {Giroletti}, {Giuliani}, {Glicenstein}, {Gnatyk},
  {Godinovic}, {Goldoni}, {G{\'o}mez-Vargas}, {Gonz{\'a}lez}, {Gonz{\'a}lez},
  {G{\"o}tz}, {Graham}, {Grandi}, {Granot}, {Green}, {Greenshaw}, {Griffiths},
  {Gunji}, {Hadasch}, {Hara}, {Hardcastle}, {Hassan}, {Hayashi}, {Hayashida},
  {Heller}, {Helo}, {Hermann}, {Hinton}, {Hnatyk}, {Hofmann}, {Holder},
  {Horan}, {H{\"o}randel}, {Horns}, {Horvath}, {Hovatta}, {Hrabovsky},
  {Hrupec}, {Humensky}, {H{\"u}tten}, {Iarlori}, {Inada}, {Inome}, {Inoue},
  {Inoue}, {Inoue}, {Iocco}, {Ioka}, {Iori}, {Ishio}, {Iwamura}, {Jamrozy},
  {Janecek}, {Jankowsky}, {Jean}, {Jung-Richardt}, {Jurysek}, {Kaaret},
  {Karkar}, {Katagiri}, {Katz}, {Kawanaka}, {Kazanas}, {Kh{\'e}lifi}, {Kieda},
  {Kimeswenger}, {Kimura}, {Kisaka}, {Knapp}, {Kn{\"o}dlseder}, {Koch},
  {Kohri}, {Komin}, {Kosack}, {Kraus}, {Krause}, {Krau{\ss}}, {Kubo}, {Kukec
  Mezek}, {Kuroda}, {Kushida}, {La Palombara}, {Lamanna}, {Lang}, {Lapington},
  {Le Blanc}, {Leach}, {Lees}, {Lefaucheur}, {Leigui de Oliveira}, {Lenain},
  {Lico}, {Limon}, {Lindfors}, {Lohse}, {Lombardi}, {Longo}, {L{\'o}pez},
  {L{\'o}pez-Coto}, {Lu}, {Lucarelli}, {Luque-Escamilla}, {Lyard}, {Maccarone},
  {Maier}, {Majumdar}, {Malaguti}, {Mandat}, {Maneva}, {Manganaro}, {Mangano},
  {Marcowith}, {Mar{\'\i}n}, {Markoff}, {Mart{\'\i}}, {Martin},
  {Mart{\'\i}nez}, {Mart{\'\i}nez}, {Masetti}, {Masuda}, {Maurin}, {Maxted},
  {Mazin}, {Medina}, {Melandri}, {Mereghetti}, {Meyer}, {Minaya}, {Mirabal},
  {Mirzoyan}, {Mitchell}, {Mizuno}, {Moderski}, {Mohammed}, {Mohrmann},
  {Montaruli}, {Moralejo}, {Morcuende-Parrilla}, {Mori}, {Morlino}, {Morris},
  {Morselli}, {Moulin}, {Mukherjee}, {Mundell}, {Murach}, {Muraishi}, {Murase},
  {Nagai}, {Nagataki}, {Nagayoshi}, {Naito}, {Nakamori}, {Nakamura}, {Niemiec},
  {Nieto}, {Niko{\l}ajuk}, {Nishijima}, {Noda}, {Nosek}, {Novosyadlyj},
  {Nozaki}, {O'Brien}, {Oakes}, {Ohira}, {Ohishi}, {Ohm}, {Okazaki}, {Okumura},
  {Ong}, {Orienti}, {Orito}, {Osborne}, {Ostrowski}, {Otte}, {Oya}, {Padovani},
  {Paizis}, {Palatiello}, {Palatka}, {Paoletti}, {Paredes}, {Pareschi},
  {Parsons}, {Pe'er}, {Pech}, {Pedaletti}, {Perri}, {Persic}, {Petrashyk},
  {Petrucci}, {Petruk}, {Peyaud}, {Pfeifer}, {Piano}, {Pisarski}, {Pita},
  {Pohl}, {Polo}, {Pozo}, {Prandini}, {Prast}, {Principe}, {Prokhorov},
  {Prokoph}, {Prouza}, {P{\"u}hlhofer}, {Punch}, {P{\"u}rckhauer}, {Queiroz},
  {Quirrenbach}, {Rain{\`o}}, {Razzaque}, {Reimer}, {Reimer}, {Reisenegger},
  {Renaud}, {Rezaeian}, {Rhode}, {Ribeiro}, {Rib{\'o}}, {Richtler}, {Rico},
  {Rieger}, {Riquelme}, {Rivoire}, {Rizi}, {Rodriguez}, {Rodriguez Fernandez},
  {Rodr{\'\i}guez V{\'a}zquez}, {Rojas}, {Romano}, {Romeo}, {Rosado}, {Rovero},
  {Rowell}, {Rudak}, {Rugliancich}, {Rulten}, {Sadeh}, {Safi-Harb}, {Saito},
  {Sakaki}, {Sakurai}, {Salina}, {S{\'a}nchez-Conde}, {Sandaker}, {Sandoval},
  {Sangiorgi}, {Sanguillon}, {Sano}, {Santander}, {Sarkar}, {Satalecka},
  {Saturni}, {Schioppa}, {Schlenstedt}, {Schneider}, {Schoorlemmer},
  {Schovanek}, {Schulz}, {Schussler}, {Schwanke}, {Sciacca}, {Scuderi},
  {Seitenzahl}, {Semikoz}, {Sergijenko}, {Servillat}, {Shalchi}, {Shellard},
  {Sidoli}, {Siejkowski}, {Sillanp{\"a}{\"a}}, {Sironi}, {Sitarek}, {Sliusar},
  {Slowikowska}, {Sol}, {Stamerra}, {Stani{\v{c}}}, {Starling}, {Stawarz},
  {Stefanik}, {Stephan}, {Stolarczyk}, {Stratta}, {Straumann}, {Suomijarvi},
  {Supanitsky}, {Tagliaferri}, {Tajima}, {Tavani}, {Tavecchio}, {Tavernet},
  {Tayabaly}, {Tejedor}, {Temnikov}, {Terada}, {Terrier}, {Terzic}, {Teshima},
  {Testa}, {Thoudam}, {Tian}, {Tibaldo}, {Tluczykont}, {Todero Peixoto},
  {Tokanai}, {Tomastik}, {Tonev}, {Tornikoski}, {Torres}, {Torresi}, {Tosti},
  {Tothill}, {Tovmassian}, {Travnicek}, {Trichard}, {Trifoglio}, {Troyano
  Pujadas}, {Tsujimoto}, {Umana}, {Vagelli}, {Vagnetti}, {Valentino},
  {Vallania}, {Valore}, {van Eldik}, {Vandenbroucke}, {Varner}, {Vasileiadis},
  {Vassiliev}, {V{\'a}zquez Acosta}, {Vecchi}, {Vega}, {Vercellone}, {Veres},
  {Vergani}, {Verzi}, {Vettolani}, {Viana}, {Vigorito}, {Villanueva}, {Voelk},
  {Vollhardt}, {Vorobiov}, {Vrastil}, {Vuillaume}, {Wagner}, {Wagner},
  {Walter}, {Ward}, {Warren}, {Watson}, {Werner}, {White}, {White},
  {Wierzcholska}, {Wilcox}, {Will}, {Williams}, {Wischnewski}, {Wood},
  {Yamamoto}, {Yamazaki}, {Yanagita}, {Yang}, {Yoshida}, {Yoshiike},
  {Yoshikoshi}, {Zacharias}, {Zaharijas}, {Zampieri}, {Zandanel}, {Zanin},
  {Zavrtanik}, {Zavrtanik}, {Zdziarski}, {Zech}, {Zechlin}, {Zhdanov},
  {Ziegler}, \& {Zorn}}]{CTA}
{Cherenkov Telescope Array Consortium}, {Acharya}, B.~S., {Agudo}, I., {et~al.}
  2019, {Science with the Cherenkov Telescope Array}, \dodoi{10.1142/10986}

\bibitem[{{Christie} {et~al.}(2019){Christie}, {Petropoulou}, {Sironi}, \&
  {Giannios}}]{Christie19}
{Christie}, I.~M., {Petropoulou}, M., {Sironi}, L., \& {Giannios}, D. 2019,
  MNRAS, 482, 65, \dodoi{10.1093/mnras/sty2636}

\bibitem[{{Deil} {et~al.}(2017){Deil}, {Zanin}, {Lefaucheur}, {Boisson},
  {Khelifi}, {Terrier}, {Wood}, {Mohrmann}, {Chakraborty}, {Watson},
  {Lopez-Coto}, {Klepser}, {Cerruti}, {Lenain}, {Acero}, {Djannati-Ata{\"\i}},
  {Pita}, {Bosnjak}, {Trichard}, {Vuillaume}, {Donath}, {Consortium}, {King},
  {Jouvin}, {Owen}, {Sipocz}, {Lennarz}, {Voruganti}, {Spir-Jacob}, {Ruiz}, \&
  {Arribas}}]{Gammapy}
{Deil}, C., {Zanin}, R., {Lefaucheur}, J., {et~al.} 2017, in International
  Cosmic Ray Conference, Vol. 301, 35th International Cosmic Ray Conference
  (ICRC2017), 766, \dodoi{10.22323/1.301.0766}

\bibitem[{{Finke} {et~al.}(2008){Finke}, {Dermer}, \& {B{\"o}ttcher}}]{Finke08}
{Finke}, J.~D., {Dermer}, C.~D., \& {B{\"o}ttcher}, M. 2008, ApJ, 686, 181,
  \dodoi{10.1086/590900}

\bibitem[{{Grolleron} {et~al.}(2023){Grolleron}, {Becerra Gonzalez}, {Biteau},
  {Cerruti}, {Grau}, \& {Gréaux}}]{Grolleron23}
{Grolleron}, G., {Becerra Gonzalez}, J., {Biteau}, J., {et~al.} 2023, in
  {proceedings of the ICRC 2023 conference, ID PGA2-35}

\bibitem[{{Wakely} \& {Horan}(2008)}]{tevcat}
{Wakely}, S.~P., \& {Horan}, D. 2008, in International Cosmic Ray Conference,
  Vol.~3, International Cosmic Ray Conference, 1341--1344

\end{thebibliography}
\bibliographystyle{aasjournal}

%
%
%

\end{document}